\documentstyle[12pt,epsfig,epsf]{article}

\def\Pom{{\bf I\!P}}

\def\lsim{\mathrel{\rlap{\lower4pt\hbox{\hskip1pt$\sim$}}
    \raise1pt\hbox{$<$}}}         

\def\gsim{\mathrel{\rlap{\lower4pt\hbox{\hskip1pt$\sim$}}
    \raise1pt\hbox{$>$}}}         

\newcommand{\be}{\begin{equation}}
\newcommand{\ee}{\end{equation}}
\newcommand{\bea}{\begin{eqnarray}}
 
\newcommand{\eea}{\end{eqnarray}}

\newcommand{\bDelta}{\mbox{\boldmath $\Delta$}}

\newcommand{\br}{{\bf r}}

\begin{document}

\begin{center}
{\bf  Diffractive DIS: Where Are We?\footnote{Invited talk 
at the International Conference on New Trends in High Eenergy Physics,
(experiment, phenomenology, theory), Yalta, Crimea, Ukraine, 
22-29 September, 2001,  }}\bigskip\\

{N.N. Nikolaev\\
{\it Institut f. Kernphysik, Forschungszentrum
J\"ulich\\ D-52425 J\"ulich, Germany\\
and\\
L.D. Landau Institute for Theoretical Physics\\ 142432 Chernogolovka, Russia}
\\ 
E-mail: N.Nikolaev@fz-juelich.de}\bigskip\\



{\bf Abstract}\\

\end{center}
A brief review of the modern QCD theory of diffractive DIS
is given. The recent progress has been remarkably rapid,
all the principal predictions from the color dipole
approach to diffraction - the $(Q^2+m_V^2)$ scaling, the 
pattern of SCHNC, shrinkage of the diffraction cone in hard 
diffractive DIS, the strong impact of longitudinal gluons 
in inclusive $J/\Psi$ production at Tevatron -, have been
confirmed experimentally.

\section{Introduction: why diffractive DIS is so fundamental}

Let us dream of e-Uranium DIS at THERA at $Q^2 \sim 10$ GeV$^2$ and $x \sim 10^{-5}$. 
Violent DIS is usually associated with complete destruction
of the target. As well known, deposition of a mere dozen MeV energy
would break the uranium nucleus entirely, but a paradoxical
yet rigorous
prediction from unitarity is that {\it diffractive} DIS $eU \to e'XU$ 
with the target nucleus emerging intact in the ground state will 
make $\approx 50\%$ of total DIS \cite{NZZnucleus}!

By the celebrated Glauber-Gribov theory of interactions
with nuclei, the abundance of diffraction
is closely related to nuclear shadowing (NS) in DIS. 
To this end we recall that inn 1974 Nikolaev 
and V.I.~Zakharov reinterpreted NS in terms of the saturation of
nuclear parton densities, caused 
by the spatial overlap of partons from different nucleons of
the Lorentz-contracted nuclei  \cite{NZ1975}. More recently, the NZ 
picture of saturation has been addressed to within QCD by Mc Lerran
and his collaborators \cite{McLerran}. If correct,
this QCD approach must inevitably lead to 50 \% diffractive DIS
- whether that would be the case or not would serve as a crucial
cross-check of the whole approach. To summarize, diffractive DIS
is a key to nuclear parton densities and QCD predictions 
for the initial state in ultrarelativistic nuclear collisions.
Because at HERA the rate of diffractive DIS is mere $10 \%$ \cite{GNZ95},
saturation effects are all but marginal, see \cite{NNZscan}.

In this talk I focus on the color dipole approach to diffraction,
a more phenomenological approach based on the Regge theory
ideas was discussed at the Conference by Kaidalov \cite{Kaidalov}.

\section{Color dipole link between inclusive
and diffractive DIS}

Structure functions (SF's) 
of DIS are related by optical theorem to the imaginary part of 
an amplitude of diagonal, $Q^2_f = Q_{in}^2 =Q^2$, forward virtual
Compton scattering (CS)
$
\gamma^{*}_{\mu}(Q^2_{in})p\to \gamma^{*}_{\nu}(Q_f^2)p'$,
which for the reason of vanishing $(\gamma^*,\gamma^*)$
momentum transfer happens to be diagonal in the photon 
helicities, ${\nu}={\mu}$. 
In the color dipole (CD) factorization \cite{NZ91} the
CS amplitude takes the form
$
A_{CS}=\Psi^{*}_{f}\otimes A_{q\bar{q}}\otimes
\Psi_{in}
$
where $\Psi_{f,in}$ is the wave function (WF) of the $q\bar{q}$ 
Fock state of the photon and the  $q\bar{q}$-proton 
scattering kernel $A_{q\bar{q}}$ is proportional to the color dipole 
cross section, which for small dipoles is related to the gluon SF 
of the target, 
$$
\sigma(x,\br)\approx {\pi^2 \over 3}r^2 \alpha_{S}({A\over r^2})
G(x,{A\over r^2})\, . 
$$
where $A\approx 10$ by properties of the Bessel functions 
\cite{NZglue}. Taking for $\Psi^{*}_{f}$ the WF of the 
vector meson (VM) or the $X=q\bar{q}$ plane waves gives the diffraction 
excitation of VM \cite{NNZscan} or hadronic continuum, 
$\gamma^* p \to Xp'$. The complete set of $q\bar{q}$ dipoles
can be substituted for by a dual complete set of vector meson states,
for the discussion of this relationship see Schildknecht's
talk at this Conference \cite{Schildknecht}. 

Alternatively, diffractive VM production 
can be obtained form CS keeping the virtuality of the initial photon, 
$Q^2=Q_{in}^2$, fixed, while continuing analytically in 
the second photon's virtuality 
to $Q_{f}^{2}=-m_{V}^2$,
$ 
\gamma^{*}_{\mu}(Q^2)p\to V_{\nu}(\bDelta)p'(-\bDelta)\, .
$ 
Diffractive VM production is accessible experimentally also at finite  $(\gamma^* V)$ momentum
transfer $\bDelta$. 

Generalization to excitation of the
$q\bar{q}g$ or higher Fock states of the photon 
is straightforward \cite{NZ92,NZ93}, within the CD approach
inclusive cross section of diffractive DIS is simply a sum
of differential cross sections of quasielastic scattering
of different Fock states of the virtual photon off
the target nucleon or nucleus. To this end, one may say
that diffractive DIS probes the partonic structure of 
the virtual photon \cite{NZ92,NZ93} in a manner closely related to 
how the structure of the deuteron is probed in
diffraction excitation of the deuteron to the proton-neutron
continuum states. With certain reservations
CD results can be reinterpreted in terms of the parton structure of
the pomeron.  For virtual photons, higher Fock states 
of the photon build up perturbatively starting from the lowest $q\bar{q}$ 
state, which entails the solid result \cite{NZ92,NZ93,GNZ95} that gluons and 
charged partons carry about an equal fraction of the momentum of the
pomeron.

The formalism set in \cite{NZ92,NZ93} and especially in 
\cite{GNZ95,GNZcharm,GNZlong,Bertini} is a basis of modern 
parameterizations of the diffractive structure function \cite{Bartels}.
 Unfortunately the use of the discredited
Ingelmann-Schlein-Regge factorization and DGLAP evolution
which is not warranted at large $\beta$ \cite{NZ93,GNZcharm,GNZlong,Bertini}.
Also, the intrinsic transverse momentum of gluons 
\cite{PomKperp} has not
yet been incorporated correctly into the diffractive
jet analysis \cite{H1DiffJet}.
For the above reasons, the still simplified form of this analysis
makes conclusions \cite{Bartels,H1DiffJet,WhitmoreCrimea} on the gluon content 
of the pomeron highly suspect.

\begin{figure}[!htb]
   \centering
   \epsfig{file=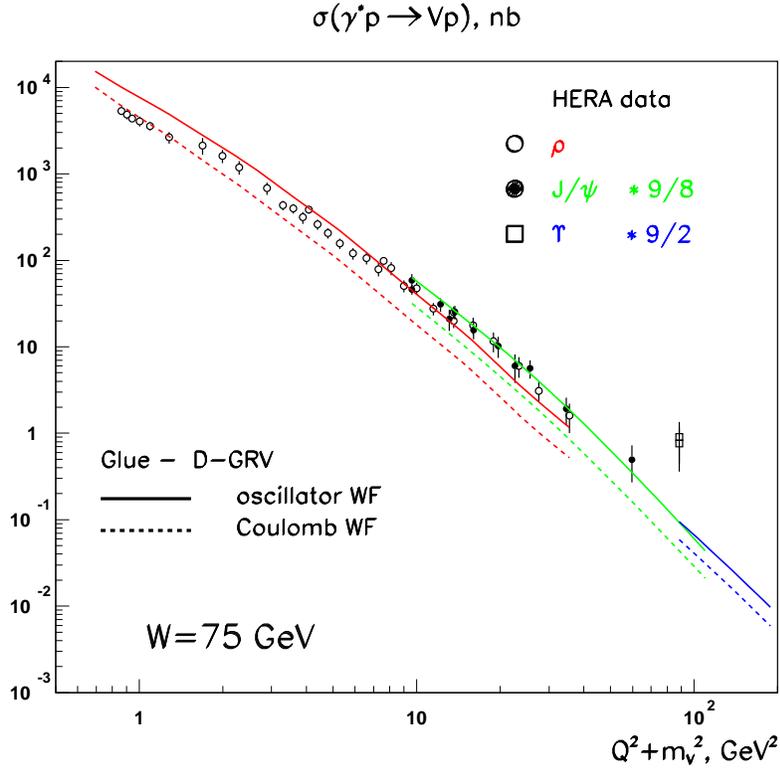,width=12cm}
\vspace{-0.5cm}
\caption{The test of the $(Q^2+m_V^2)$ scaling. The divergence of the 
solid and dashed curves indicates the sensitivity to the WF of the VM.
The experimental data are from HERA \protect{\cite{HERAdata}}.} 
\label{Figure1}
\end{figure}

\section{The $Q^2+m_V^2$ scaling}

While  DIS probes CD
$\sigma(x,\br)$ in a broad
range of ${1 \over AQ^2} \lsim r^2 \lsim 1 $ fm$^2$,
the diffractive VM 
production probes $\sigma(x,\br)$ at a 
scanning radius  \cite{NNNComments,NNZscan} 
$$
r\sim r_{S}= {6\over \sqrt{Q^2 + m_{V}^2}}\, ,
$$
and the gluon SF of the target at the hard scale
$\overline{Q}^2 \approx$ (0.1-0.25)$* (Q^2 + m_{V}^2)$ 
and $x=0.5(Q^2+m_{V}^2)/(Q^2+W^2)$. After factoring out the 
charge-isospin factors, that entails the $(Q^2 + m_{V}^2)$ scaling 
of the VM production cross section\cite{NNZscan}, see fig.~1. The
same scaling holds also
for the effective intercept $\alpha_{\Pom}(0)-1$ of the energy
dependence of the production amplitude and
contribution to the diffraction slope $B$ from the $\gamma^* \to V$
transition vertex, which is $\propto r_{S}^2$ and exhibits the 
$(Q^2 + m_{V}^2)$ scaling \cite{NZZslope,JETPslope}, see fig.~2.
This $(Q^2 + m_{V}^2)$ scaling formulated in 1974, has recently become
a popular way of presenting the experimental data \cite{WhitmoreCrimea}.
The theoretical calculations \cite{Igor} are based on the differential 
glue in the proton found in \cite{INdiffglue}

\begin{figure}[!htb]
\vspace{-0.5cm}   
\centering
   \epsfig{file=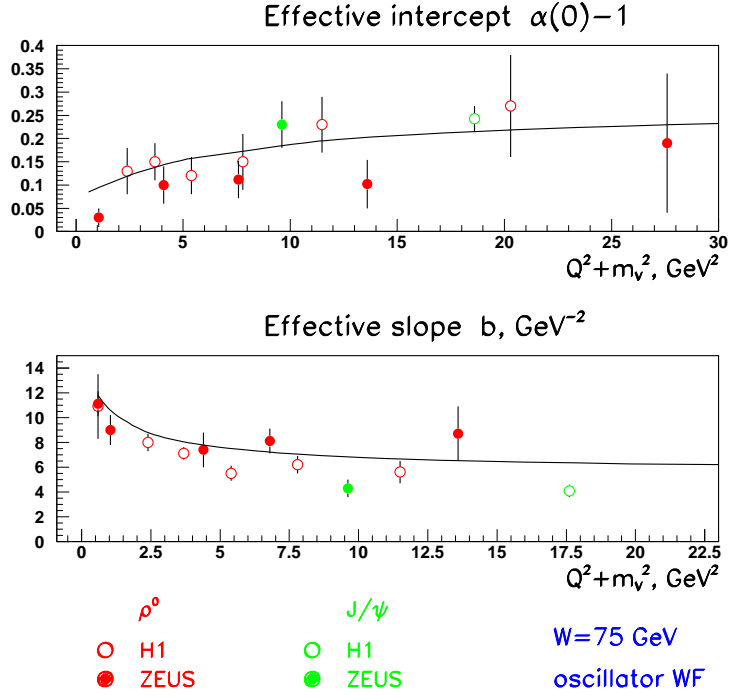,width=11cm}
\vspace{-0.5cm}
\caption{The $(Q^2+m_V^2)$ scaling of the effective intercept and 
diffraction slope \protect{\cite{Igor}}}
\label{Figure2}
\end{figure}

\section{Shrinkage of the diffraction cone in hard diffraction}

Gribov's shrinkage of the diffraction cone 
$B = B_{0}+2\alpha'_{\Pom}\log W^2$, quantized in terms of
the slope  $\alpha'_{\Pom}$ of the pomeron trajectory, is the
salient feature of hadronic scattering.
By the unitarity relation, diffractive elastic scattering is
driven by mutiproduction processes, and in the unitarity context
the shrinkage of the diffraction cone is well known to derive from 
the Gribov-Feinberg-Chernavski diffusion in the
impact parameter space. 

At this Conference, we heard from Whitmore of the ZEUS finding
\cite{WhitmoreCrimea}
of the shrinkage of the diffraction cone in $\gamma p\to J/\Psi p$
with the result 
$$
\alpha'_{eff} =0.122\pm 0.033(stat)+0.018-0.032(syst)~~
GeV^{-2}\, .
$$ 
Precisely such
a shrinkage has been predicted in 1995 by Nikolaev, Zakharov and
Zoller \cite{NZZslope} to persist within QCD even for
hard processes. 

First, in the usual approximation $\alpha_S=const$
the BFKL pomeron is the fixed cut in the complex-$j$ plane \cite{FKL}. 
Already in their first, 1975,  publication on QCD pomeron, Kuraev,
Lipatov and Fadin commented that incorporation of the asymptotic
freedom splits the cut into a sequence of  moving Regge poles \cite{FKL}, 
see Lipatov \cite{Lipatov} for more details.  
Within the CD approach, the Regge trajectories
of these poles where calculated in \cite{NZZslope,NZZpoles}. 
The CD cross section satisfies the CD BFKL equation \cite{NZZBFKL},
$$
\partial \sigma(x,r)/\partial\log{1\over x} = {\cal K}\otimes  \sigma(x,r)\, ,
$$
which has the Regge solutions 
$$\sigma_{n}(x,r) =\sigma_{n}(r)x^{-\Delta_{n}}.$$
The CD kernel ${\cal K}$ is related to the flux of Weizs\"acker-Williams gluons
around the $q\bar{q}$ dipole. 

The NZZ strategy was to evaluate $\alpha_{n}'$ from the energy dependence
of $\lambda(x,r) = B\sigma(x,r)$, which satisfies the
inhomogeneous equation 
$$\partial \lambda(x,r)/\partial\log{1\over x} - 
{\cal K}\otimes  
\lambda(x,r) = {\cal L}\otimes r^2 \sigma(x,r),$$ 
and has solutions
$$
\lambda_{n}(x,r) =2\sigma_{n}(x,r)\cdot \alpha'_{n} \log{1\over x}\, ,$$
which correspond to the Regge rise of the diffraction slope
with energy.
Because $B= {1\over 2}
\langle b^2\rangle $
and the impact parameter $b$ receives a contribution from the gluon-$q\bar{q}$
separation $\rho$, the inhomogeneous term 
${\cal L}\otimes r^2 \sigma(x,r)$ is driven by precisely 
the impact parameter diffusion of WW gluons. Because 
$\alpha'_{n}$ is driven by the inhomogeneous term, there is 
a manifest relationship between $\alpha'_n$ and the 
Gribov-Feinberg-Chernavski diffusion in the
impact parameter space. Evidently, the dimensionfull quantity $\alpha'_{n}$
depends on the infrared regularization of QCD, within the specific
regularization \cite{NZZBFKL,NZHERA} which has lead to an extremely successful
description of the proton structure function (see \cite{NSZpion} and references therein),
for the rightmost hard BFKL pole we found $\alpha'_{\Pom}\approx 0.07$ GeV$^-2$.
The contribution from subleading BFKL poles was found to be still 
substantial at subasymptotic energy of HERA with the result
$\alpha_{eff}'\approx 0.15-0.17$ GeV$^{-2}$ \cite{NZZslope}, this
prediction from 1995 agrees perfectly with the ZEUS finding.

\section{Digression: Longinudinal photons and gluons in DIS}

In DIS incident leptons serve as 
a source of virtual photons and experimentally one studies a
virtual photoproduction of various hadronic states. 
While real photons are 
transverse ones, i.e., have only circular 
polarizations, $\mu = \pm 1$,  
virtual photons radiated by leptons have also the longitudinal 
polarization, which in the 
scaling limit equals  
\be
\epsilon_L = {2(1-y) \over 2(1-y)+y^2},
\label{eq:1.3}
\ee
where $y$ is a fraction of the beam lepton energy taken away by the photon,
so that the photoabsorption cross section measured in 
the inclusive DIS equals
$\sigma = \sigma_T+\epsilon_L \sigma_L$. The effect  of 
longitudinal photons,
quantified by $R_{DIS}=\sigma_L/\sigma_T \sim 0.2$, is marginal, though.

The branching of gluons into gluons is a dominant feature of QCD evolution
at small $x$. In the conventional collinear approximation one
treats gluons as having only the physical transverse polarizations.
However, in close analogy to virtual photons, virtual gluons 
have also a substantial longitudinal polarization. In striking
 contrast 
to {\sl inclusive} DIS, diffractive excitation of VM and 
small mass continuum
is that they are entirely dominated 
by $\sigma_{L}$ \cite{NNZscan,GNZlong}. As we shall see below,
in inclusive production too interaction of longitudinal virtual
photons could be outstanding in defiance of the collinear
factorization \cite{JPsiPsiPrim}.

\section{Spin dependence of vector meson production}

Regarding the spin dependence of diffractive VM, the fundamental
point 
is that the sum of quark and antiquark helicities equals helicity of 
neither the photon nor vector meson. If for the 
nonrelativistic massive quarks, $m_{f}^{2} \gg Q^{2}$ the only allowed 
transition is $\gamma^{*}_{\mu} \to q_{\lambda} +\bar{q}_{\bar{\lambda}}$ 
with $\lambda +\bar{\lambda}=\mu$. In the relativistic case transitions
of transverse photons $\gamma^{*}_{\pm}$ into the $q\bar{q}$ state with 
$\lambda +\bar{\lambda}=0$, 
in which the helicity of the photon is transferred to the $q\bar{q}$ orbital 
 momentum, are equally allowed. Consequently, in QCD the 
$s$-channel helicity non-conserving (SCHNC) transitions
$$\gamma^{*}_{\pm} \to (q\bar{q})_{\lambda +\bar{\lambda}=0} \to
\gamma^{*}_{L}$$
and
$$
\gamma^{*}_{\pm} \to (q\bar{q})_{\lambda +\bar{\lambda}=0 }\to
\gamma^{*}_{\mp} $$ 
are allowed \cite{NPZLT,KNZ98} and SCHNC persists at small $x$
despite the exact conservation of the helicity of quarks in  
$q\bar{q}$-target scattering. This argument 
for SCHNC does not require the applicability of pQCD.
Furthermore, the leading  contribution to the proton structure function
comes entirely from SCHNC transitions of transverse photons
\cite{INdiffglue} - the
fact never mentioned in textbooks.

We emphasize that SCHNC helicity flip only is possible due to 
the transverse and/or longitudinal Fermi motion of quarks and
is extremely sensitive to spin-orbit coupling in the
vector meson, I refer for details to \cite{KNZ98,INDwave}.
The consistent analysis of production of 
$S$-wave and $D$-wave vector mesons is presented only
in \cite{INDwave}.  The dominant 
SCHNC effect in vector meson production is the interference 
of SCHC $\gamma^{*}_{L} \to V_L$ and SCHNC $\gamma^{*}_{T}\to V_L$ production,
i.e., the element $r_{00}^{5}$ of the vector meson spin
density matrix. The overall agreement
between our theoretical estimates 
\cite{Igor} of the 
spin density matrix $r_{ik}^{n}$ for
diffractive  $\rho^{0}$ assuming pure $S$-wave in the $\rho^{0}$-meson
and the ZEUS \cite{ZEUSflip} and H1 \cite{H1flip} experimental data
is very good, there is a clear evidence for  $r_{00}^{5}\neq 0$, see
fig.~3.

\begin{figure}[!htb]
   \centering
   \epsfig{file=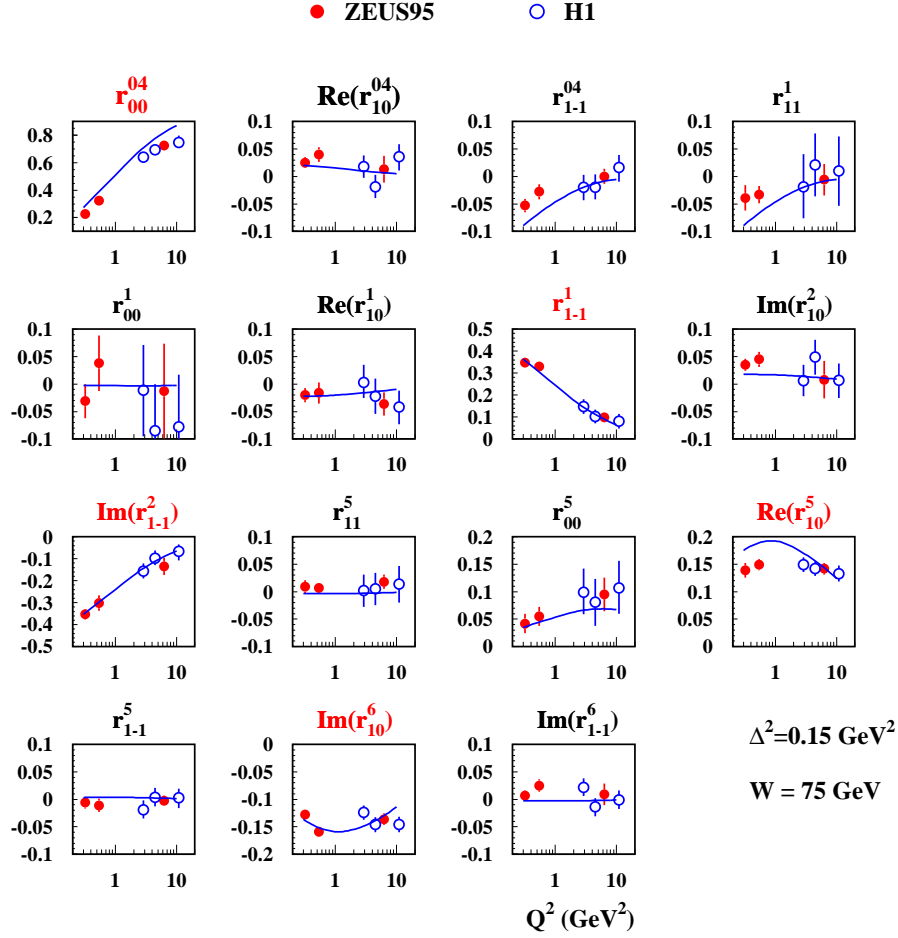,width=13cm}
\vspace{-0.5cm}
\caption{Predictions for the spin density matrix in the $\rho^{0}$
production vs. the experimental data from HERA \protect{\cite{ZEUSflip,H1flip}}.}
\label{Figure3}
\end{figure}

\begin{figure}[!htb]
\vspace{-.3cm}
   \centering
   \epsfig{file=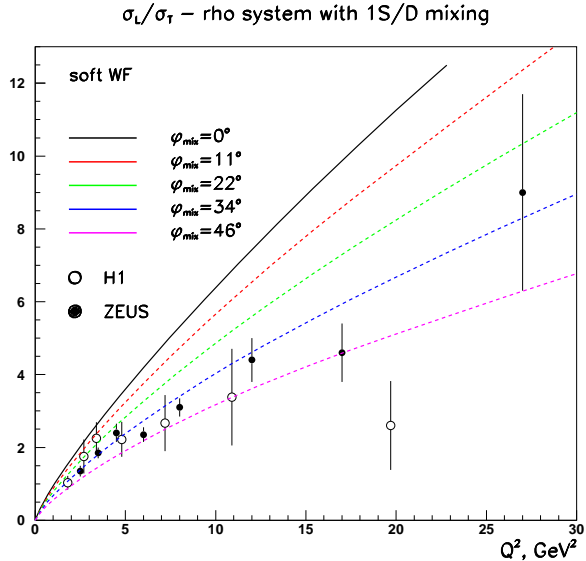,width=9cm}
\vspace{-0.5cm}
\caption{The sensitivity of $R=\sigma_L/\sigma_T$ for
the $\rho$ production to the S-D-mixing \protect{\cite{Igor}},
for the compilation of the experimental data 
see \protect{\cite{ZEUSLT}}.}
\label{Figure4}
\vspace{-.3cm}
\end{figure}

\section{Issues with $R=\sigma_L/\sigma_T$}

Still another fundamental point about spin properties of
diffractive DIS is that the vertex of the SCHC 
transition
$\gamma^*_L \to (q\bar{q})_{\lambda +\bar{\lambda}=0}$ is proptional
to $Q$, which entails \cite{NNZscan}
$$
R= {\sigma_L(\gamma^*_L p\to V_L p) \over 
\sigma_T(\gamma^*_T p\to V_T p)} \sim { Q^2 \over m_V^2} >> 1
$$
for diffractive VM's. As was first noticed 
in \cite{NNZscan}, a numerical analysis with realistic 
soft WF gives values of $R$ substantially smaller than a crude
estimate $R \approx Q^2/m_{V}^2$. Still, the theoretical 
calculations \cite{Igor} seem to overpredict 
$R=\sigma_L/\sigma_T$  at large $Q^2$, see fig.~4,
for the compilation of the experimental data see \cite{ZEUSLT}. 

As it was shown in \cite{INDwave}, $R=\sigma_L/\sigma_T$ is
very different for the $S$ and $D$-wave states. As a result,
an admissible
$S-D$ mixing brings the theory to a better agreement
with the data, see fig.~4. Furthermore, the recent data from ZEUS 
\cite{ZEUSLT}
do indicate, the experimental value of $R$ tends to rise with
the time. 
Here I would like to raise the issue of sensitivity of
$R$ to the short distance 
properties of vector mesons \cite{NNNCracow}. 

Consider $R_{el}= \sigma_{L}/\sigma_T$ for elastic CS 
$\gamma^*p\to \gamma^*p$, which is quadratic in the ratio of
CS amplitudes. By optical theorem one finds
$$
R_{el}= {\sigma_L(\gamma^*_L p\to \gamma^*_L p) \over 
\sigma_T(\gamma^*_T p\to \gamma^*_T p)}= \left|{A(\gamma^*_L p\to \gamma^*_L p) \over 
A(\gamma^*_T p\to \gamma^*_T p)}\right|^2= 
\left({\sigma_{L}\over \sigma_T}\right)^2_{DIS} \approx 4\cdot 10^{-2}
$$
Here I used the prediction \cite{NZHERA} for inclusive DIS
$R_{DIS} = \left.\sigma_{L}/ \sigma_T\right|_{DIS}\approx 0.2$, which
is consistent with the indirect experimental evaluations of
$R_{DIS}$ at HERA. 
Such a dramatic change from $R_{el} \ll 1$ to $R \sim Q^2/m_V^2
>> 1$
suggests that predictions fir $R$ in diffractive  VM
production are extremely sensitive to the poorly known
admixture of quasi-pointlike $q\bar{q}$ componets in VM. 

\section{Longitudinal gluons and polarization of a
direct $J/\Psi$ and $\Psi'$ at Tevatron}

There is a long standing mystery of the predominant  
longitudinal polarization of prompt $J\Psi$ and $\Psi'$
produced at large transverse momentum $p_{\perp}$ as
observed by the CDF collaboration in
inclusive $p\bar{p}$ interactions at Tevatron
\cite{CDF}, see fig.~5, in which the polarization
parameter'
$$
\alpha= {\sigma_T -2\sigma_L \over \sigma_T+\sigma_L}
$$
is shown (the observed $\Psi$' are arguably the direct ones,
the prompt $J/\Psi$'s include the  $J/\Psi$'s both
from the direct production and decays of higher charmonium 
states) . Specifically, the color-octet model
\cite{BraatenSigma} is able to parameterize the 
observed reaction cross section
(for a criticism of the standard formulation of 
the color-octet model see \cite{Regensburg}), 
but fails badly
in its predictions \cite{Leibovich,BraatenPolar} 
for the polarization parameter $\alpha$, see fig.~5.

\begin{figure}[!htb]
   \centering
   \epsfig{file=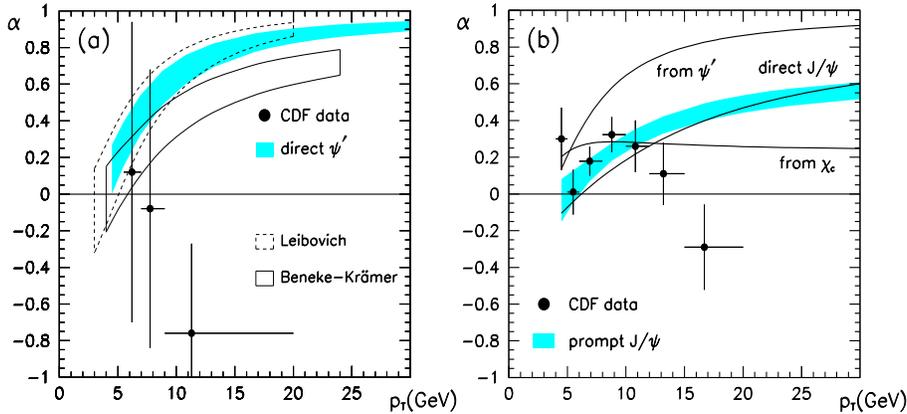,width=12cm}
\vspace{-0.3cm}
\caption{Predictions \protect{\cite{Leibovich,BraatenPolar}}
 from the color-octet 
model for the polarization parameter $\alpha$ vs. $p_{\perp}$ for
direct $\Psi$ and prompt $J/\Psi$ compared to CDF data
\protect{\cite{CDF}}.}
\label{Figure5}
\end{figure}

Arguably, production of charmonium states at mid-rapidity 
is controlled by gluon-gluon collisions. Now recall that 
in the standard collinear factorization the colliding 
gluons are regarded as on-mass shell, and transversely
polarized, ones. Which is the principal reason, why one
predicts the predominantly transverse polarization of
the produced $J/\Psi$ and $\Psi$'.

The QCD subprocesses for direct production of $C=-1$ vector 
states of charmonium are shown in fig.~6. In order to emphasize
an impact of their virtuality of the colliding gluons, I
indicate explicitly the origin of the gluon $g^{*}$. As
I mentioned in section 5, beside the familiar
transverse polarization, highly virtual gluons have also the
hitherto ignored longitudinal polarization
\cite{JPsiPsiPrim}.

\begin{figure}[!htb]
   \centering
   \epsfig{file=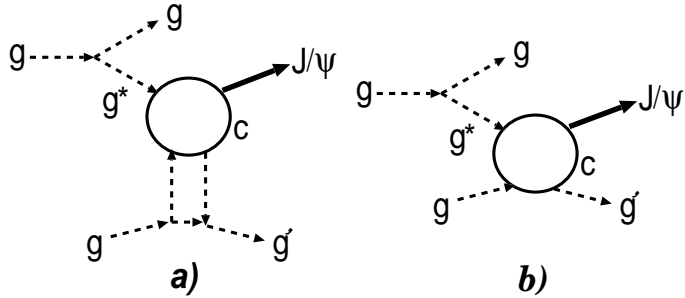,width=9cm}
\vspace{-0.3cm}
\caption{The diffractive QCD subprocesses for
the production of prompt vector states of
charmonium in hadronic reactions.}
\label{Figure6}
\vspace{-.3cm}
\end{figure}

In order to illustrate our principal point let me focus
on the "diffractive" mechanism of fig.~6a. 
It dominates at large invariant masses 
$\hat{w}$ of the $Vg'$ system, $\hat{w}^2 \gg M_{\Psi}^2$.
The virtuality of the gluon $g^*$ is
controlled by the transverse momentum $k_{\perp}$ of
the gluon $g^*$, so that $Q^2 \approx k_{\perp}^2$. 
The sub-process $g^* + g \to J/\Psi g'$
proceeds predominantly in the forward direction, which
implies that the transverse momentum of the $g^*$ is
transferred to the $J/\Psi$, so that
$p_{\perp} \approx k_{\perp}$.  The difference between color octet
two-gluon state in the $t$-channel of fig.~6a and
color-singlet two gluon state in the diffractive
pomeron exchange is completely irrelevant for spin
properties of the $J/\Psi$ production, and for the diffractive 
mechanism of fig.~6a we
unequivocally predict
$R =\sigma_L/\sigma_T \sim p_{\perp}^2/m_{\Psi}^2$,
i.e., $\alpha \to -1$ for very large $p_{\perp}$.
After some color algebra, one can readily relate the 
total cross section of the "diffractive" mechanism to
the cross section of photoproduction of $J/\Psi$ on
nucleons. We found that "diffractive" mechanism is
short of strength and could explain only $\sim 10$ per
cent of the observed yield of the direct $J/\Psi$. 

The diagrams of fig.~6b dominate for $\hat{w} \sim m_{\Psi}$.
Arguably, the above estimate for the $p_{\perp}$ dependence
of $R$ applied to this mechanism too. Crude estimates 
show that the contribution from this mechanism is commensurate 
to, or larger than, that from the "diffractive" large-$\hat{w}$
mechanism.

Besides the predominantly "forward" production when the
transverse momentum of the $g^*$ is transferred predominantly
to the direct $J/\Psi$, one must also consider the 
large angle reaction $g^* g \to J/\Psi + g$, which could affect
the polarization parameter $\alpha$. The full 
numerical analysis has not been completed yet, still 
we believe that the so far neglected longitudinal gluons
resolve a riddle of the longitudinal
polarization of direct $J\Psi$ and $\Psi'$.

\section{Conclusions}

The QCD theory of diffractive DIS is gradually coming of age.
The fundamental $(Q^2+m_V^2)$ scaling predicted in 1994 has
finally been recognized by experimentalists. 
The shrinkage of the diffraction cone for hard photoproduction
$\gamma p \to J/\Psi p$ discovered by the ZEUS collaboration
is the single most important result. It shows that the BFKL
pomeron is a (series of) moving pole(s) in the complex-$j$
plane. The slope of the pomeron trajectory and the rate of
shrinkage of the diffraction cone for hard photo- and
electroproduction predicted in 1995 has been confirmed
experimentally. 
So far neglected
longitudinal gluons are predicted to dominate production
of direct vector mesons at large transverse momentum
in hadronic collisions and resolve the long-standing
riddle of the dominant longitudinal polarization
of the $J/\Psi$ and $\Psi'$ discovered by CDF.

Thanks are due to my collaborators Igor Ivanov and
Wolfgang Sch\"afer for much insight and pleasure of
joint work on the ideas reported here. 
I'm grateful to Prof. L. Jenkovszky for the invitation to 
Crimea-2001. This work was partly supported by
the INTAS grants 97-30494 and 00-00366.

\end{document}